\documentclass[aps,prb,preprint,groupedaddress,showpacs,amsmath,amssymb]{revtex4}

\usepackage{graphicx}
\usepackage{dcolumn}
\usepackage{bm}

\begin{document}

\title{Thermal expansion and anisotropic pressure derivatives of $T_c$ in Ba(Fe$_{1-x}$Co$_x$)$_2$As$_2$ single crystals}

\author{S. L. Bud'ko,$^1$ N. Ni,$^1$ S. Nandi,$^1$ G. M. Schmiedeshoff,$^2$ and P. C. Canfield$^1$}
\affiliation{$^1$Ames Laboratory US DOE and Department of Physics and Astronomy, Iowa State University, Ames, IA 50011, USA
\\
$^2$Department of Physics, Occidental College, Los Angeles, CA 90041, USA}

\date{\today}

\begin{abstract}
Heat capacity and anisotropic thermal expansion was measured for Ba(Fe$_{1-x}$Co$_x$)$_2$As$_2$ ($x = 0, 0.038, 0.074$) single crystals. Thermal expansion is anisotropic and, in tetragonal phase, is significantly higher along the $c$-axis.  Previously reported phase transitions, including possibly split structural and magnetic for $x = 0.038$ are clearly seen in both measurements. Uniaxial pressure derivatives of the superconducting transition temperature inferred from the Ehrenfest relation have opposite signs for in-plane and $c$-axis pressures for both Ba(Fe$_{0.962}$Co$_{0.038}$)$_2$As$_2$ and Ba(Fe$_{0.926}$Co$_{0.074}$)$_2$As$_2$, with the opposite sign of this anisotropy.

\end{abstract}


\maketitle

\section{Introduction}

The discovery of superconductivity in F-doped LaFeAsO \cite{kam08a} and K-doped BaFe$_2$As$_2$ \cite{rot08a} compounds caused an outpouring of experimental and theoretical studies of the materials containing Fe-As layers as a structural unit. The understanding that Co substitution for Fe in the (AE)Fe$_2$As$_2$ (AE = Ba, Sr, Ca) could be used to both stabilize superconductivity \cite{sef08a,lei08a,kum08a} and simplify growth while improving homogeneity, makes the (AE)(Fe$_{1-x}$Co$_x$)$_2$As$_2$ series (and in particular Ba(Fe$_{1-x}$Co$_x$)$_2$As$_2$ \cite{sef08a,ahi08a,nin08a,chu08a,gor08a}) a model system for studies of physical properties of Fe-As based materials.

The $T - x$ phase diagram for Ba(Fe$_{1-x}$Co$_x$)$_2$As$_2$ is rather complex: \cite{nin08a,chu08a,gor08a} on Co-doping the critical temperature of the structural/antiferromagnetic transition decreases, possibly with a separation in critical temperatures between these two transitions, \cite{nin08a,chu08a} then, above $x \sim 0.035$ superconductivity is observed, apparently in the orthorhombic/antiferromagnetic phase. Above $x \sim 0.06$ the structural/magnetic transitions vanish, whereas superconductivity appears to persist to $x \sim 0.12$. In this work we report temperature dependent heat capacity and thermal expansion for three samples of Ba(Fe$_{1-x}$Co$_x$)$_2$As$_2$ with Co concentration chosen so that they are located in three different parts of the $x - T$ phase diagram: pure BaFe$_2$As$_2$ to serve as a reference point, Ba(Fe$_{0.962}$Co$_{0.038}$)$_2$As$_2$ for which superconductivity in orthorhombic/antiferromagnetic phase is reported, \cite{nin08a,chu08a,gor08a} and Ba(Fe$_{0.926}$Co$_{0.074}$)$_2$As$_2$ that is located close to the maximum of the superconducting dome and for which no structural/magnetic phase transition was reported.

Thermal expansion is uniquely sensitive to magnetic, structural and superconducting transitions \cite{kro98a}. In addition, the combination of specific heat and thermal expansion data, in the case of the second order phase transitions (the superconducting transitions in the Ba(Fe$_{1-x}$Co$_2$)$_2$As$_2$  series), allows for the evaluation of the uniaxial, often anisotropic, pressure derivatives of the critical temperatures and possibly get some insight in structure - physical properties relations in the series.

\section{Experimental Methods}

Single crystals of Ba(Fe$_{1-x}$Co$_x$)$_2$As$_2$ were grown out of self flux using conventional high-temperature solution growth techniques. \cite{sef08a,can92a} Detailed description of the crystal growth procedure for this series can be found elsewhere. \cite{nin08a} The samples are plate-like with the plates being perpendicular to the crystallographic $c$-axis. The in-plane orientation of the samples at room temperature was determined by the x-ray Laue technique. Elemental analysis of the samples was performed using wavelength dispersive x-ray spectroscopy (WDS) in the electron probe microanalyzer of a JEOL JXA-8200 electronmicroprobe. \cite{nin08a} Measured rather than nominal Co concentrations will be used throughout the text. The heat capacity of the samples was measured using a hybrid adiabatic relaxation technique of the heat capacity option in a Quantum Design, PPMS-14 instrument. Thermal expansion data were obtained using a capacitive dilatometer constructed of OFHC copper; a detailed description of the dilatometer is presented elsewhere \cite{sch06a}. The dilatometer was mounted in a Quantum Design PPMS-14 instrument and was operated over a temperature range of 1.8 to 305 K. The samples were cut and lightly polished so as to have parallel surfaces perpendicular to the $[100]$ and $[001]$ directions with the distances $L$ between the surfaces ranging between approximately $0.25 - 3$ mm. Care was taken to remove small parts of the samples exfoliated near the edges, so that the measurements were performed on the bulk part of the crystals.  Specific heat and thermal expansion data were taken on warming.

\section{Results and Discussion}

\begin{figure}
\begin{center}
\includegraphics[angle=0,width=120mm]{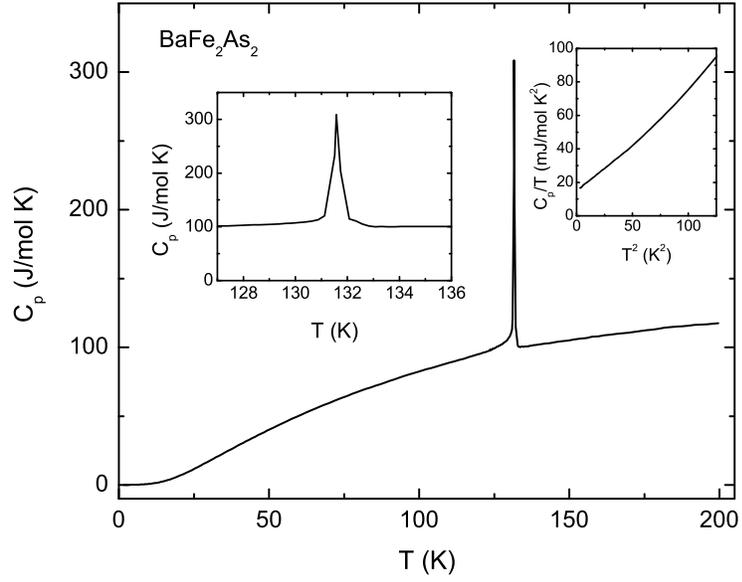}
\end{center}
\caption{Temperature-dependent heat capacity of BaFe$_2$As$_2$ single crystal. Left inset: enlarged region near the structural/magnetic phase transition; right inset: low temperature $C_p/T$ as a function of $T^2$. Linear specific heat term, $\gamma$ is $\approx 15$ mJ/mol K$^2$.}\label{F1}
\end{figure}

\begin{figure}
\begin{center}
\includegraphics[angle=0,width=90mm]{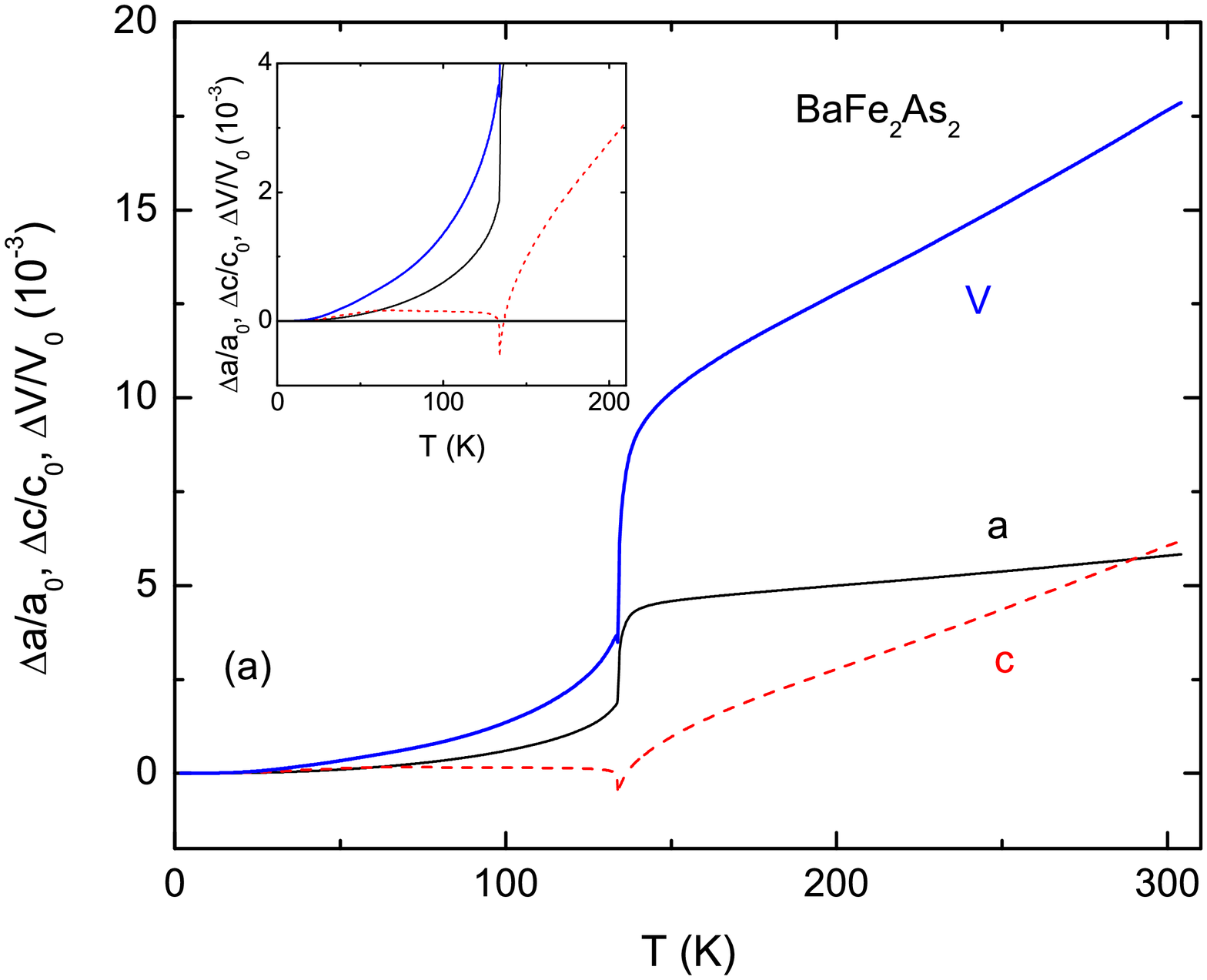}
\includegraphics[angle=0,width=90mm]{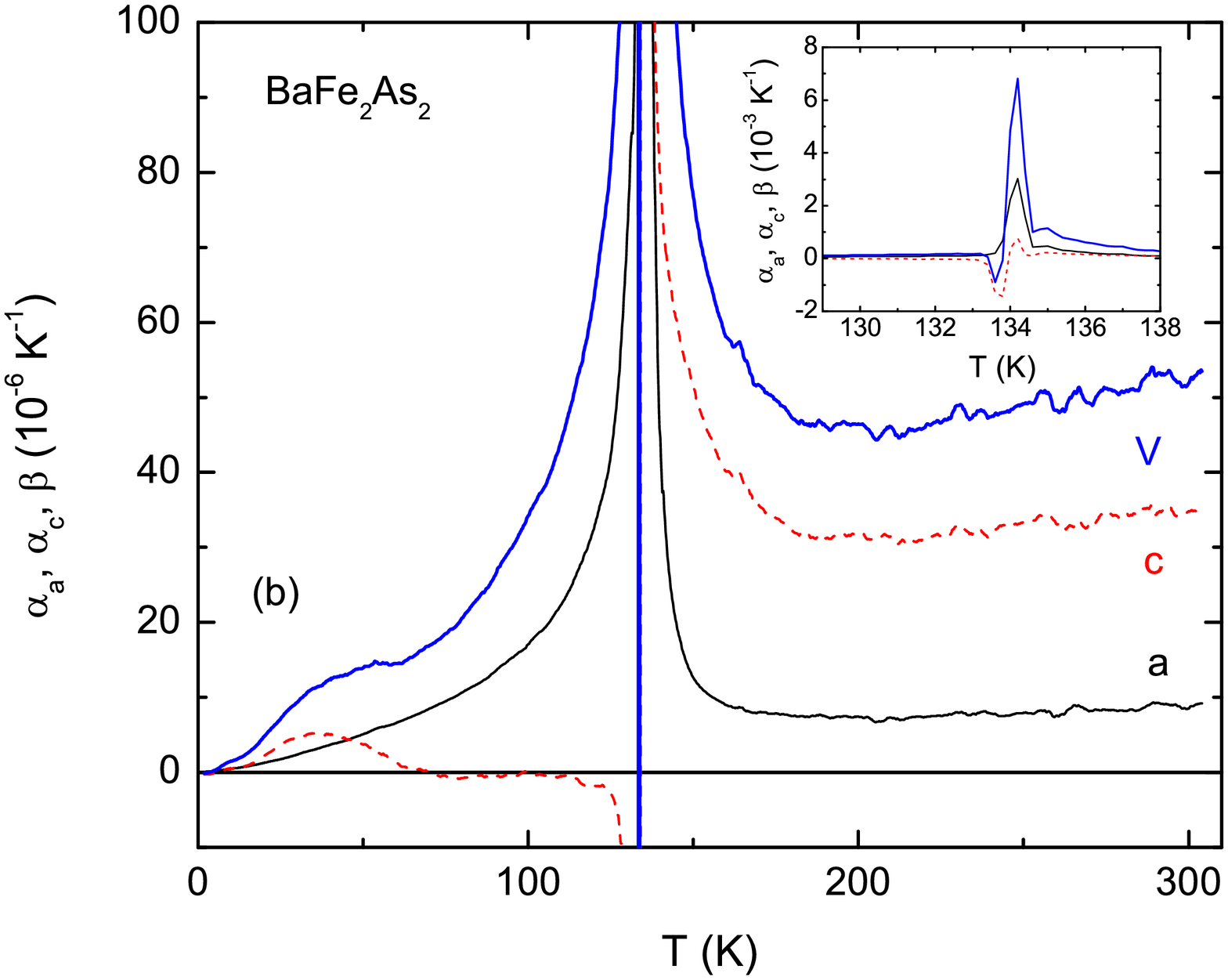}
\end{center}
\caption{(Color online)  Anisotropic, temperature dependent,(a) linear and volume thermal expansivities ($\Delta a/a_0, \Delta c/c_0$ and $\Delta V/V_0$, where the changes are taken relative to the values at 1.8 K) and (b) thermal expansion coefficients ($\alpha_a, \alpha_c$ and $\beta$) of  BaFe$_2$As$_2$ single crystal. Insert to (a): expanded low temperature part of the main graph. Insert to (b): expanded part of the main graph near the structural/magnetic phase transition. Note that at temperatures below the structural/magnetic phase transition the data marked as $a$-axis are some average of $a$ and $b$.}\label{F2}
\end{figure}

The heat capacity of the pure BaFe$_2$As$_2$ is shown in Fig. \ref{F1}. The feature associated with the structural/magnetic transition is narrow and sharp. The transition temperature is slightly lower than that reported for polycrystalline samples \cite{rot08b,hua08a} but consistent with the results on similarly grown single crystals \cite{chu08a}. The extreme sensitivity of the structural/magnetic transition temperature e.g. to small amount of impurities was clearly seen on crystals grown out of Sn flux. \cite{nin08b} The value of the electronic specific heat coefficient can be estimated as $\gamma \approx 15$ mJ/mol K$^2$ (right inset to Fig. \ref{F1}), in an agreement with the result in [\onlinecite{rot08b}]. Anisotropic, temperature dependent, linear and volume thermal expansivities ($\Delta a/a_0, \Delta c/c_0$ and $\Delta V/V_0$, where the changes are taken relative to the values at 1.8 K) and thermal expansion coefficients ($\alpha_a, \alpha_c$ and $\beta$) are shown in Fig. \ref{F2}. The structural/magnetic phase transition is clearly seen in the data. Above the transition, the relative change in the $c$-axis thermal expansivity is significantly larger than that in the $a$-axis (Fig. \ref{F2}(a)) a fact that is reflected in more than factor of 3 higher thermal expansion coefficient in this temperature region (Fig. \ref{F2}(b)). The thermal expansion coefficients are almost temperature-independent between $\sim 200 - 300$ K. At low temperatures, $\alpha_c$ is close to zero at $\sim 60 - 110$ K.

Thermal expansion in BaFe$_2$As$_2$ (for the sample grown out of Sn flux and with structural/magnetic transition at $\sim 90$ K) was measured via neutron diffraction. \cite{suy08a} In the tetragonal phase, near room temperature, the $\alpha_c$ is similar to our data, whereas $\alpha_a$ and $\beta$ are reported to be smaller than in our measurements. This difference may be due to the difference in samples as well as may reflect the differences in accuracy of the two experimental techniques.
\\

\begin{figure}
\begin{center}
\includegraphics[angle=0,width=120mm]{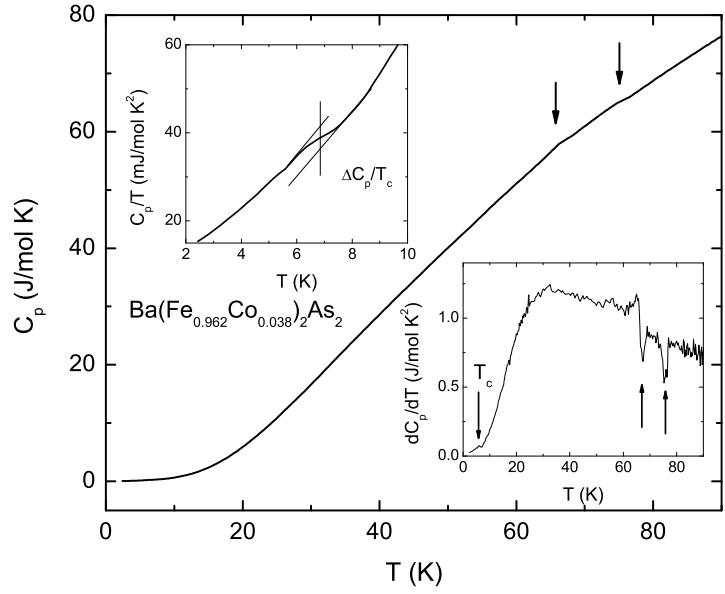}
\end{center}
\caption{Temperature-dependent heat capacity of Ba(Fe$_{0.962}$Co$_{0.038}$)$_2$As$_2$ single crystal. Insets: $C_p/T$ vs. $T$ near the superconducting transition with the estimate of $\Delta C_p/T_c$ shown; temperature dependence of the derivative $dC_p/dT$. Arrows mark the phase transitions.}\label{F3}
\end{figure}

\begin{figure}
\begin{center}
\includegraphics[angle=0,width=80mm]{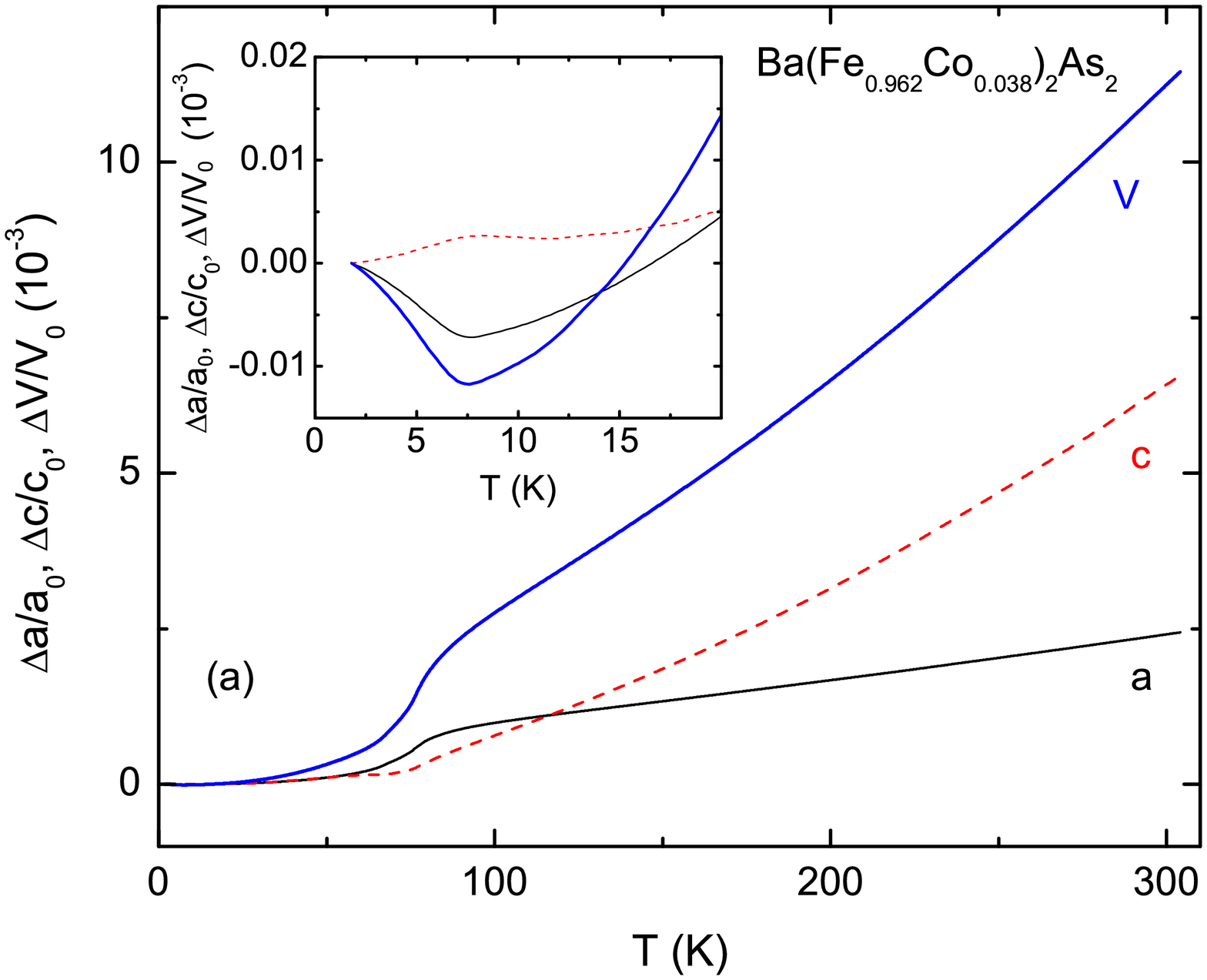}
\includegraphics[angle=0,width=80mm]{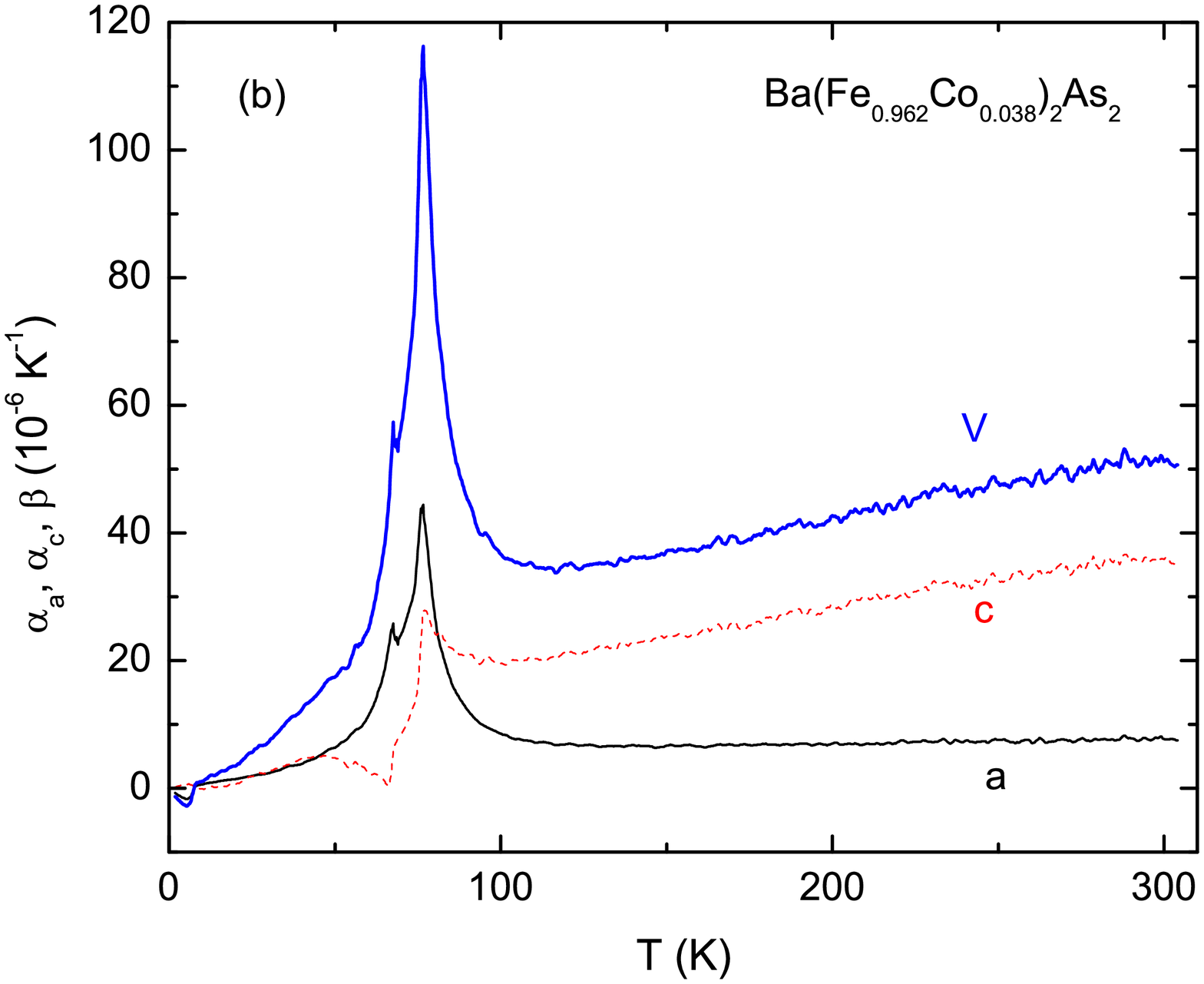}
\includegraphics[angle=0,width=80mm]{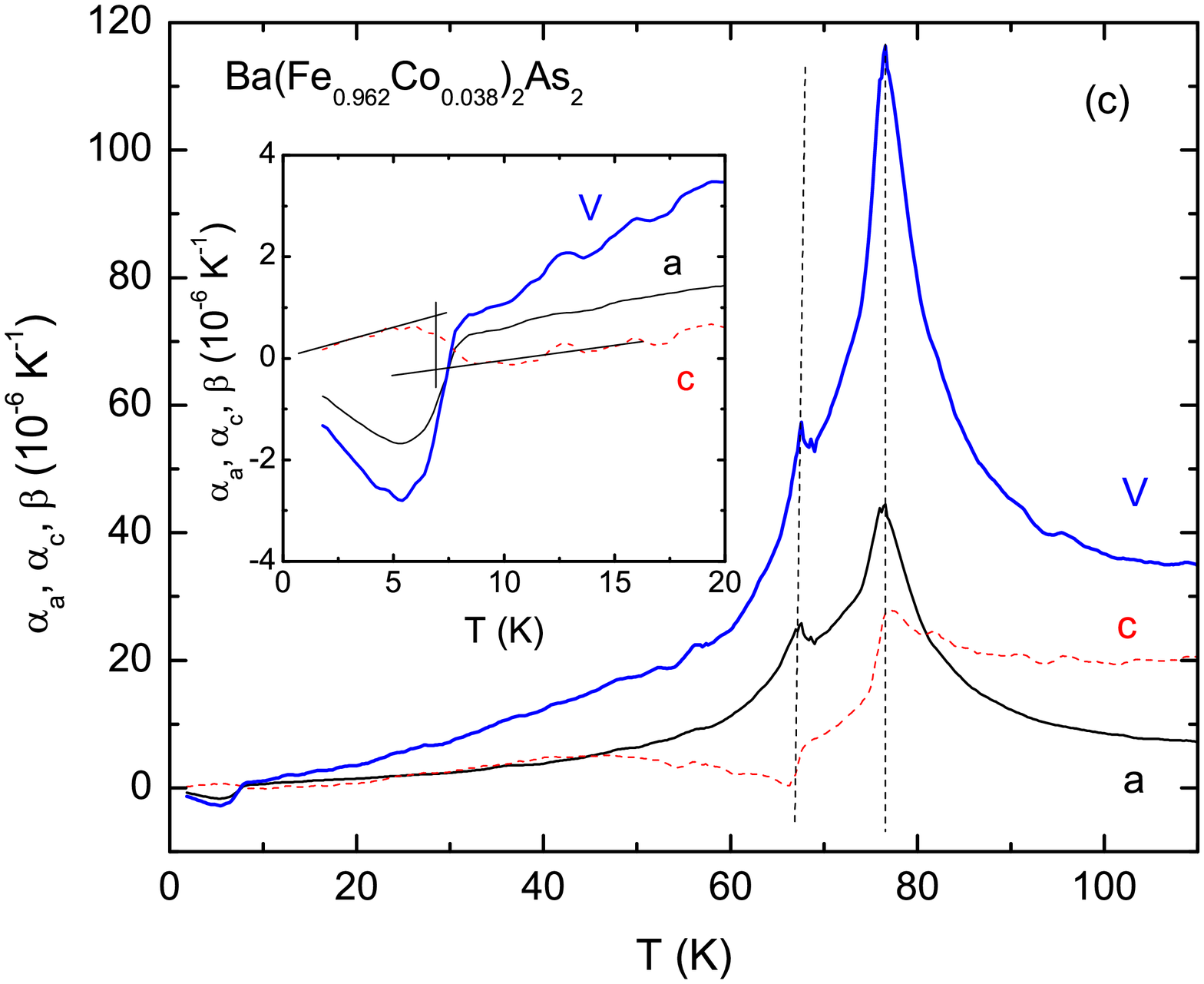}
\end{center}
\caption{(Color online)  (a) Linear and volume thermal expansivities (the changes are taken relative to the values at 1.8 K) of  Ba(Fe$_{0.962}$Co$_{0.038}$)$_2$As$_2$ single crystal; inset: enlarged low temperature part of the data. (b) Thermal expansion coefficients of Ba(Fe$_{0.962}$Co$_{0.038}$)$_2$As$_2$ single crystal. (c) Enlarged, low temperature, part of the data in the panel (b), vertical lines mark structural/magnetic phase transitions; inset: enlarged low temperature part of the data including the superconducting transition, lines show how $\Delta \alpha_c$ was estimated.  Note that at temperatures below the structural/magnetic phase transitions the data marked as $a$-axis are some average of $a$ and $b$.}\label{F4}
\end{figure}

The temperature dependent heat capacity of Ba(Fe$_{0.962}$Co$_{0.038}$)$_2$As$_2$ [\onlinecite{nin08a}] is shown in Fig. \ref{F3}. Two features, separated by $\sim 9$ K are seen above 60 K. These features are possibly the result of a separation between the structural and magnetic transitions with Co doping \cite{nin08a,chu08a}, although the hypothesis of the chemical inhomogeneity of the sample \cite{nin08a} cannot be readily excluded. The jump in the specific heat at the superconducting transition can be estimated as $\Delta C_p/T_c \approx 4$ mJ/mol K$^2$. Using the BCS weak coupling approximation, $\Delta C_p/\gamma T_c = 1.43$ and assuming 100\% superconducting fraction, we can estimate the normal state electronic specific heat coefficient, $\gamma$ for Ba(Fe$_{0.962}$Co$_{0.038}$)$_2$As$_2$ as $\sim 2.8$ mJ/mol K$^2$. (The fact that this is a relatively low value will be discussed in detail below.) Anisotropic thermal expansivities and thermal expansion coefficients are shown in Fig. \ref{F4}. Near room temperature the thermal expansion values of Ba(Fe$_{0.962}$Co$_{0.038}$)$_2$As$_2$ are similar to that of BaFe$_2$As$_2$.  For Ba(Fe$_{0.962}$Co$_{0.038}$)$_2$As$_2$ all phase transitions detected in the heat capacity are clearly seen at similar temperatures in the thermal expansion data. As in the $C_p(T)$ data, two distinct features are observed above 60 K. The higher temperature feature is stronger and if these two anomalies are associated with the split structural and magnetic transitions (as opposed to coinciding as in BaFe$_2$As$_2$) it seems reasonable to assume that the more pronounced, upper, transition is structural and the lower one is magnetic, because thermal expansion more directly couples to structural rather than magnetic transition for which a sizable magnetoelastic term is required.

Anomalies corresponding to the superconducting transition are unambiguous in the thermal expansion data (Fig. \ref{F4}(c), inset). The signs of $\Delta \alpha_a$ and $\Delta \alpha_c$ at the superconducting transition are different. The initial uniaxial pressure derivatives of the superconducting transition temperature, $T_c$, can be estimated using the Ehrenfest relation \cite{bar99a} for the second order phase transitions:
\begin{displaymath}
dT_c/dp_i = \frac{V_m \Delta\alpha_i T_c}{\Delta C_p}
\end{displaymath}
where $V_m$ is the molar volume ($V_m \approx 6.14~10^{-5}$ m$^3$ for Ba(Fe$_{0.962}$Co$_{0.038}$)$_2$As$_2$ [\onlinecite{nin08a}]), $\Delta\alpha_i$ is a change of the linear ($i = a, c$) or volume ($\alpha_V = \beta$) thermal expansion coefficient at the phase transition, and $\Delta C_p$ is a change of the specific heat at the phase transition. The inferred (via the Ehrenfest relation) uniaxial pressure derivatives are listed in Table \ref{T1}. It is noteworthy, that from these results in-plane uniaxial pressure apparently should cause a decrease of $T_c$, whereas $T_c$ is expected to increase when the uniaxial pressure is applied along the $c$-axis. For a "hydrostatic" pressure the uniaxial effects are partially compensated.
\\

\begin{figure}
\begin{center}
\includegraphics[angle=0,width=120mm]{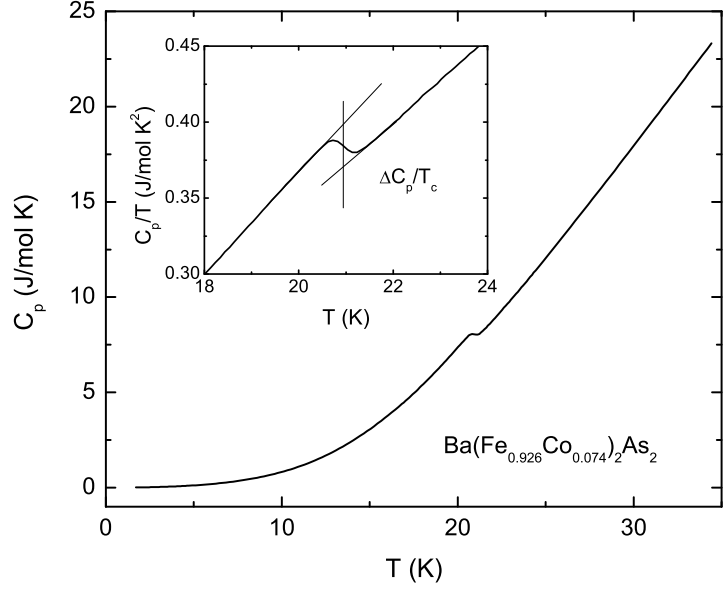}
\end{center}
\caption{Temperature-dependent heat capacity of Ba(Fe$_{0.926}$Co$_{0.074}$)$_2$As$_2$ single crystal. Insets: $C_p/T$ vs. $T$ near the superconducting transition with the estimate of $\Delta C_p/T_c$ shown.}\label{F5}
\end{figure}

The temperature dependent heat capacity of Ba(Fe$_{0.926}$Co$_{0.074}$)$_2$As$_2$ [\onlinecite{nin08a}] is shown in Fig. \ref{F5}. The jump in the specific heat at the superconducting transition can be estimated as $\Delta C_p/T_c \approx 28$ mJ/mol K$^2$. Using the BCS weak coupling approximation, $\Delta C_p/\gamma T_c = 1.43$ and assuming 100\% superconducting fraction, we can estimate the normal state electronic specific heat coefficient, $\gamma$ for Ba(Fe$_{0.926}$Co$_{0.074 }$)$_2$As$_2$ as $\sim 19.6$ mJ/mol K$^2$. There may be several reasons for the significant difference in values of the electronic specific heat coefficient, $\gamma$, inferred from $\Delta C_p/T_c$: (i) density of states may change on Co - doping \cite{sef08a}, however the extreme sensitivity of band structure calculations e.g. to the position of As in the structure precludes accurate evaluation of this contribution; (ii) in principle, it is possible the value of $\Delta C_p/\gamma T_c$ changes with Co doping and differs from the BCS weak coupling value 1.43, however this contribution is not expected to be larger than factor of $\sim 2$; \cite{car90a} (iii) we cannot exclude that inhomogeneities, e.g. of Co concentration, exist in Ba(Fe$_{0.962}$Co$_{0.038}$)$_2$As$_2$ samples \cite{nin08a} and cause significant underestimate of the $\Delta C_p/\gamma T_c$ value for this material (due to broadening of the transition and, possibly, less than 100\% superconducting fraction). Additional effort is required to address this question.

\begin{figure}
\begin{center}
\includegraphics[angle=0,width=90mm]{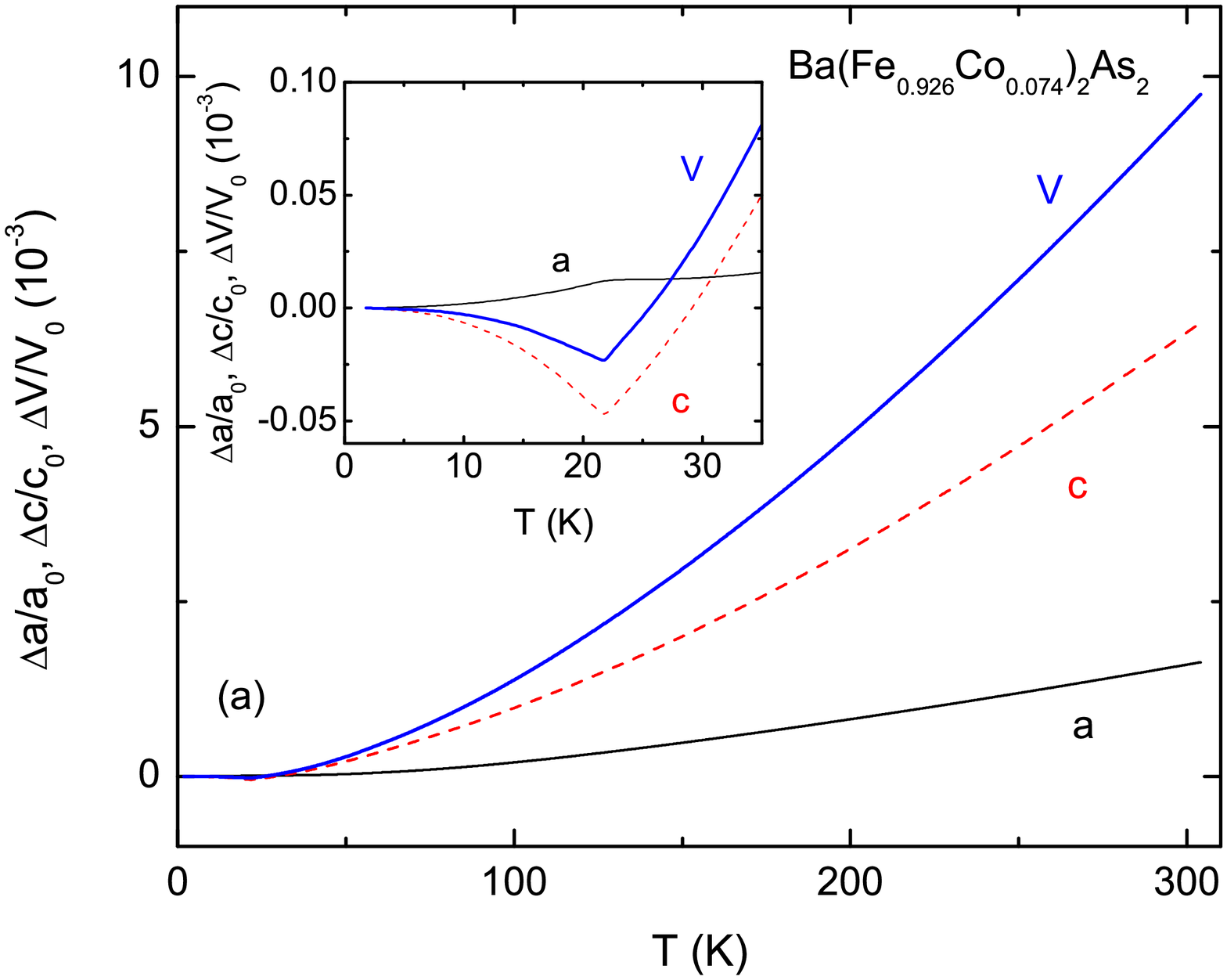}
\includegraphics[angle=0,width=90mm]{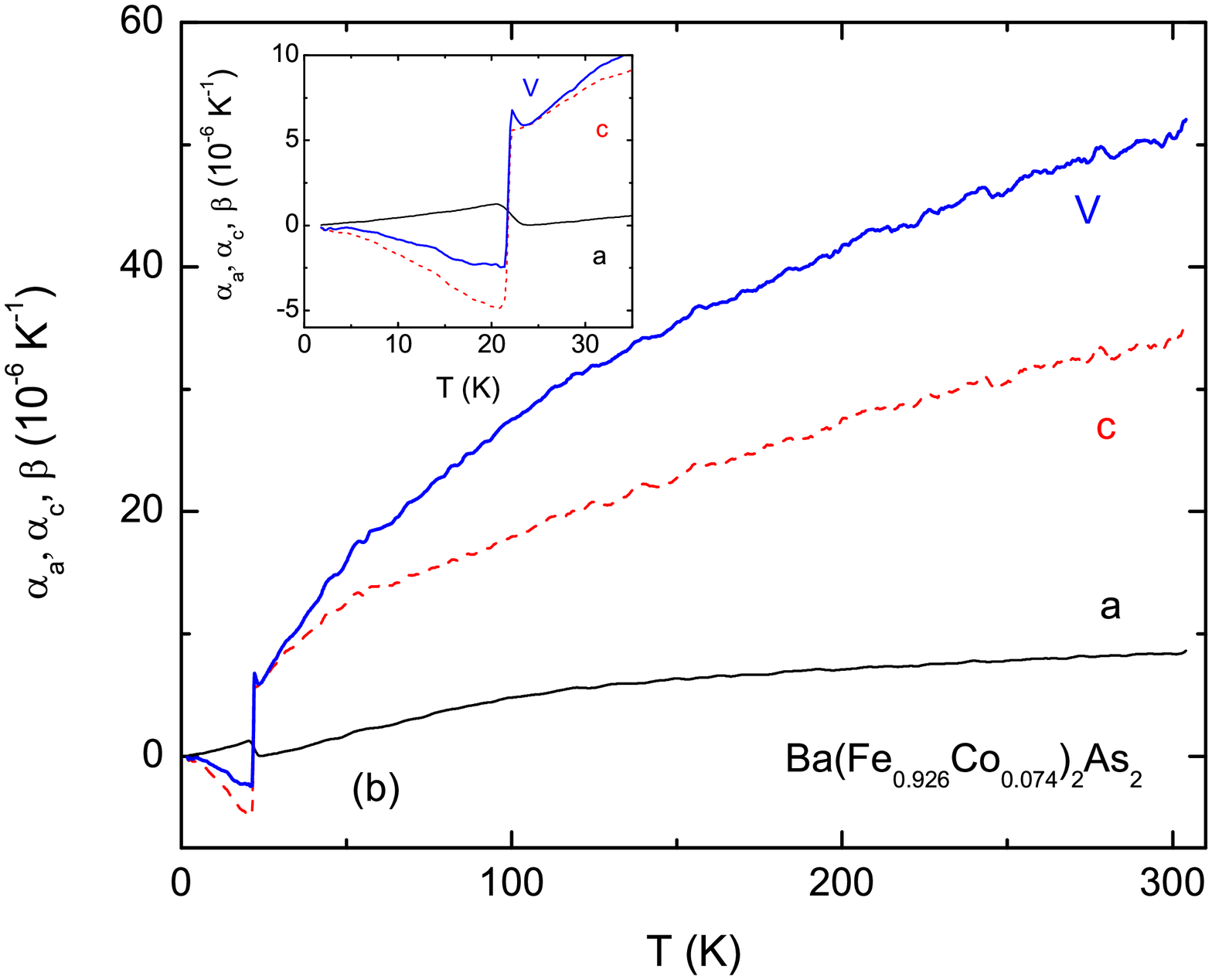}
\end{center}
\caption{(Color online)  (a) Linear and volume thermal expansivities (the changes are taken relative to the values at 1.8 K) of  Ba(Fe$_{0.926}$Co$_{0.074}$)$_2$As$_2$ single crystal; inset: enlarged low temperature part of the data. (b) Thermal expansion coefficients of Ba(Fe$_{0.926}$Co$_{0.074}$)$_2$As$_2$ single crystal; inset: enlarged low temperature part of the data including the superconducting transition.}\label{F6}
\end{figure}

Anisotropic thermal expansivities and thermal expansion coefficients for  Ba(Fe$_{0.926}$Co$_{0.074}$)$_2$As$_2$ are shown in Fig. \ref{F6}. Near room temperature the thermal expansion values are similar to that of BaFe$_2$As$_2$ and Ba(Fe$_{0.962}$Co$_{0.038}$)$_2$As$_2$. Anomalies corresponding to the superconducting transition are clearly seen in the thermal expansion data (Fig. \ref{F6}(b)). The signs of $\Delta \alpha_a$ and $\Delta \alpha_c$ at superconducting transition are different and opposite to that of Ba(Fe$_{0.962}$Co$_{0.038}$)$_2$As$_2$. Using the heat capacity data and the Ehrenfest relation, for Ba(Fe$_{0.926}$Co$_{0.074}$)$_2$As$_2$ $T_c$ is expected to increase under in-plane uniaxial pressure and to decrease when the uniaxial pressure is applied along the $c$-axis. Again, for a "hydrostatic" pressure the uniaxial effects are partially compensated. The initial uniaxial pressure derivatives of  $T_c$ estimated using the Ehrenfest relation are listed in Table \ref{T1}. These results can be compared with a recent study of the anisotropic thermal expansion of the Co-doped sample with a similar doping level, Ba(Fe$_{0.92}$Co$_{0.08}$)$_2$As$_2$ [\onlinecite{har08a}]. Room temperature thermal expansion coefficients are similar to our data, as well as the sign and the order of magnitude of $\Delta \alpha_a$ and $\Delta \alpha_c$ at superconducting transition. The $\Delta \alpha_a$ values are very close in both measurements, whereas $\Delta \alpha_c$, measured along the shorted crystal dimension, are somewhat different, possibly reflecting the difference between the samples and errors in the measurements. Still the main result, different signs of the in-plane and $c$-axis uniaxial pressure derivatives, is clear and consistent in both measurements.
\\

The inferred "hydrostatic" pressure derivatives (Table \ref{T1}) can be (semi-qualitatively) compared with the results of direct, resistance measurements under pressure \cite{ahi08a} on Ba(Fe$_{1-x}$Co$_x$)$_2$As$_2$ with nominal Co concentrations $x = 0.04$ and 0.10 ($dT_c/dP = 0.4$ K/kbar and 0.065 K/kbar respectively). The comparison is rather poor, with the discrepancies possibly caused by the difficulties in thermal expansion measurements, in particular along the $c$-axis, due to morphology of the samples, as well as possible contribution from Poisson stresses in these soft anisotropic materials.
\\

The opposite signs of the uniaxial pressure derivatives of $T_c$ were observed in other materials, both "conventional" \cite{bud06a,bud06b} and high temperature \cite{gug94a} superconductors. Moreover, for YBa$_2$Cu$_3$O$_{7-\delta}$ even in-plane uniaxial pressure derivatives, $dT_c/dp_a$ and $dT_c/dp_b$ have different signs. \cite{mei91a,bud92a,wel92a,wel94a} However, at least for La$_{2-x}$Sr$_x$CuO$_4$,  in which $T_c(x)$ dependence has similar dome shape, the sign of the uniaxial pressure derivatives was reported independent of doping. \cite{gug94a}.

\section{Summary}

Thermal expansion of Ba(Fe$_{1-x}$Co$_x$)$_2$As$_2$ ($x = 0, 0.038$ and 0.074) single crystals is anisotropic, with $\alpha_c \sim 3~\alpha_a$ at room temperature, in the tetragonal phase. The values of $\alpha_a$ and $\alpha_c$ at room temperature are very similar for all three Co-concentrations. Features associated with the structural/magnetic ($x = 0, 0.038$) and superconducting ($x = 0.038, 0.074$) transitions are clearly seen both in heat capacity and thermal expansion data. For $x  = 0.038$ two features above 60 K are observed in both measurements, with the higher temperature anomaly more pronounced than the lower one, in thermal expansion, that suggests that the higher anomaly is associated with the structural tansition and the lower one with the magnetic. Uniaxial pressure derivatives of $T_c$ are evaluated, based on the Ehrenfest relation.  For Ba(Fe$_{0.962}$Co$_{0.038}$)$_2$As$_2$ $dT_c/dp_a < 0$ and $dT_c/dp_c > 0$, as opposed to $dT_c/dp_a > 0$ and $dT_c/dp_c < 0$ for Ba(Fe$_{0.926}$Co$_{0.074}$)$_2$As$_2$. So for both samples have anisotropic uniaxial pressure derivatives of $T_c$ but the sign of the anisotropy is different. This is an unexpected result and it requires confirmation and understanding. It is consistent though with the change in low temperature crystal structure from orthorhombic for Ba(Fe$_{0.962}$Co$_{0.038}$)$_2$As$_2$ to tetragonal for Ba(Fe$_{0.926}$Co$_{0.074}$)$_2$As$_2$ effecting signature of the superconducting transition. This result serves as a further evidence of the superconducting dome existing in both orthorhombic and tetragonal regions.

\begin{acknowledgments}

Work at the Ames Laboratory was supported by the US Department of Energy - Basic Energy Sciences under Contract No. DE-AC02-07CH11358. GMS was supported by the National Science Foundation under grant DMR-0704406. GMS acknowledges encouragement from Prof. V. Q. Heffalump. We thank Andreas Kreyssig  for useful discussions and help in Laue orientation of the crystals and Jiaqiang Yan for help in synthesis.

\end{acknowledgments}

\clearpage

\begin{table}

\begin{tabular}{|c||c|c|c|c|c|c|c|c|}
\hline
$x$-Co&$\Delta C_p/T_c$&$\Delta \alpha_a(T_c)$&$\Delta \alpha_c(T_c)$&$\Delta \beta(T_c)$&$dT_c/dp_a$&$dT_c/dp_c$&$dT_c/dP$\\
&mJ/mol K$^2$&10$^{-6}$ K$^{-1}$&10$^{-6}$ K$^{-1}$&10$^{-6}$ K$^{-1}$&K/kbar&K/kbar&K/kbar\\
\hline
0.038&4&-2.7&1.1&-4.0&-4.1&1.7&-6.1\\
\hline
0.074&28&1.4&-11.7&-8.1&0.3&-2.6&-1.8\\
\hline
\end{tabular}
\caption{Uniaxial pressure derivatives of superconducting transition temperature in Ba(Fe$_{0.962}$Co$_{0.038}$)$_2$As$_2$  and Ba(Fe$_{0.926}$Co$_{0.074}$)$_2$As$_2$ ($dT_c/dP$ is calculated as $2 \cdot dT_c/dp_a + dT_c/dp_c$).}
\label{T1}
\end{table}

\clearpage

\clearpage

\clearpage

\clearpage

\clearpage

\clearpage

\end{document}